\documentclass[twocolumn,superscriptaddress,nofootinbib]{revtex4}

\usepackage{amssymb}
\usepackage{amsmath}
\usepackage{multirow}
\usepackage{mathrsfs}

\usepackage{graphicx}
\usepackage{color}
\usepackage{dcolumn}
\usepackage{bm}
\newfont{\Fr}{eufm10}
\usepackage{hyperref}

\begin{document}

\title{Inhomogeneities from quantum collapse scheme without inflation}
\author{Gabriel R. Bengochea}
\email{gabriel@iafe.uba.ar} \affiliation{Instituto de Astronom\'\i
a y F\'\i sica del Espacio (IAFE), UBA-CONICET, CC 67, Suc. 28,
1428 Buenos Aires, Argentina}
\author{Pedro \surname{Ca\~nate}}
\email{pedro.canate@nucleares.unam.mx}
\affiliation{Instituto de Ciencias Nucleares, UNAM \\
  M\'exico D.F. 04510, M\'exico}
\author{Daniel \surname{Sudarsky}}
\email{sudarsky@nucleares.unam.mx}
\affiliation{Instituto de Ciencias Nucleares, UNAM \\
  M\'exico D.F. 04510, M\'exico}

\begin{abstract}

In this work, we consider the problem of the emergence of seeds of
cosmic structure in the framework of the non-inflationary model
proposed by Hollands and Wald. In particular, we consider a
modification to that proposal designed to account for breaking the
symmetries of the initial quantum state, leading to the generation
of the primordial inhomogeneities. This new ingredient is
described in terms of a spontaneous reduction of the wave
function. We investigate under which conditions one can recover an
essentially scale free spectrum of primordial inhomogeneities, and
which are the dominant deviations that arise in the model as a
consequence of the introduction of the collapse of the quantum
state into that scenario.

\end{abstract}

\pacs{Valid PACS appear here}

\keywords{Cosmology, Inflation, Non-inflationary models, Quantum
collapse schemes} \maketitle

\section{Introduction}

Inflation is presently considered as an integral part of our
understanding of cosmological evolution. Inflationary models are
generally credited with explaining the large scale homogeneity,
isotropy and flatness of our universe as well as accounting for
the origin of the seeds of cosmic structure. All structures in our
universe emerge from a featureless stage described by a background
Friedmann-Robertson-Walker (FRW) cosmology with a nearly
exponential expansion driven by the potential of a single scalar
field, and from its quantum fluctuations characterized by a simple
vacuum state.

In inflationary models, the modes relevant to cosmological
perturbations are assumed to be born in their ground state at a
time when their proper wavelengths are much shorter than the
Hubble radius. That is, the state of the quantum field
characterizing the seeds of structure is determined by the
instantaneous vacuum state corresponding to the static universe
that would be obtained by freezing the cosmological evolution in
very early epochs (a precise description of such quantum state
construction for arbitrary space times can be seen for instance in
\cite{WALD-QFTCS}). The resulting state in the case of a later
exponential expansion (which is close to what is given in simple
inflationary models) is known as the Bunch-Davies vacuum.

In the Heisenberg picture, the state of the field at later times
is, of course, unchanged and the evolution of the field operators
is encoded in the evolution of the modes. From such a setting, one
obtains a fluctuation spectrum for these modes which corresponds
to a scale free spectrum, and thus fits very well with the
Harrison-Zel'dovich prediction \cite{H-Z}. Furthermore, that
primordial spectrum, upon incorporation of plasma physics effects,
taking place before decoupling, leads to a prediction for the
Cosmic Microwave Background (CMB) that fits the data extremely
well \cite{wmap9, planck1}.

In a recent work \cite{wald}, Hollands and Wald proposed an
alternative model involving a simple fluid, which, under some more
or less natural hypothesis, would be equally capable to reproduce
the scale free spectrum and thus, in principle, account for the
seeds of cosmic structure. Therefore, it may not be necessary to
assume that an era of inflation actually occurred and to postulate
the existence of a new fundamental scalar such as the inflaton
field. The natural hypothesis referred to above concern the
'birth time' and the initial state of the relevant modes, and will
be discussed in more detail below.

In that work, Hollands and Wald also argue that the resolution of
the flatness and horizon problems provided by inflation are not
truly satisfactory, and that only a much deeper understanding of
the conditions determining the initial state could shed light on
that matter. We do not wish to engage in this part of the
discussion here.

The model put forward in \cite{wald}, starts with the assumption
that semiclassical physics applies to phenomena on spatial scales
larger than some fundamental length $l_0$, which, presumably, is
of order of the Planck length ($l_p$) or the grand unification
scale. Following this point of view, the authors argue, it would
be natural to treat the modes as effectively being born at a time
when their proper wavelength is equal to the fundamental scale
$l_0$. Consequently, all the modes would be continuously created
over all time. Assuming that the modes are created in their
instantaneous ground state (a state that is thus isotropic and
homogeneous), the authors obtained a prediction for a scale free
spectrum with appropriate amplitude.

The starting point of the analysis corresponds to a universe that
is well described by a flat FRW metric, and where there is a large
background stress-energy that is linearly perturbed by quantum
fluctuations. According to the model, the matter in the early
universe can be described, on spatial scales greater than $l_0$,
by a fluid with equation of state $p=w\:\rho$, where $w$ is a
constant. The analysis then continues along a line close to that
followed in the standard analysis of inflationary models, by
quantizing the perturbations of the coupled Einstein-fluid system,
according to Section 10.2 of \cite{muk}.

Finally, the resulting power spectrum is obtained (as is
customary in this kind of analysis), by considering the two point
function in the above described state of the quantum field. As
noted, this result is such that for modes with wavelength greater
than the Hubble radius, at the time of decoupling, one finds a
scale free spectrum of density perturbations. The adequate value
for the spectrum amplitude is obtained by adjusting the value for
$l_0$, ($l_p/l_0 \sim 10^{-5}$). Furthermore, the authors argue
that, within this approach, one can expect a larger ratio of
tensor to scalar perturbations than that which are typical for the
inflationary models. We plan to study this issue in a future work.

In the present manuscript, we focus on the following issue: it is
a fact that in both, the inflationary picture as in the proposal
by Hollands and Wald, one must face the difficulties posed when
considering a quantum description for the early universe. A form
that the problem takes in the present setting is that, according
to the two proposals, a completely homogeneous and isotropic stage
(evolving according to dynamical equations that cannot break such
symmetries), must nevertheless lead, after some time, to a
universe containing actual inhomogeneities and anisotropies,
presumably characterized by the fluctuation spectrum. This issue
has been considered at length in other works, including detailed
discussions of the shortcomings of the most popular attempts to
address the problem, and we will not repeat such extensive
discussions here, except for a brief description intended only as
an introduction for the reader who is not familiar with the
problem. It is clear that such transition from a symmetric
situation to one that is not, cannot be simply the result of
quantum unitary evolution, since, as we noted, the dynamics does
not break these initial symmetries of the system. As discussed in
\cite{sud}, and despite multiple claims to the contrary (e.g.
\cite{kiefer}), there is no satisfactory solution to this problem
within the standard physical paradigms. Recently, some books that
presents the standard inflationary  paradigm  have   referred to
this subject  acknowledging  to  a certain  extent  the unresolved
difficulty (see e.g. \cite{weinberg, mukbook, lyth, penbook}).

The proposal to handle this shortcoming was considered for the
first time in \cite{pss}. There, the problem was addressed by
introducing a new ingredient into the inflationary account of the
origin of the seeds of cosmic structure: the self-induced collapse
hypothesis. The basic idea is that an internally induced
spontaneous collapse of the wave function of the inflaton field is
the mechanism by which inhomogeneities and anisotropies arise at
each particular scale. That proposal was inspired on early ones
for the resolution of the measurement problem in quantum theory
\cite{Bhom-Bub, Pearle:76, Pearle:79, PCS:13, GRW:85, GRW:86,
Pearle:89, GRW:90}, which regarded the collapse of the wave
function as an actual physical process taking place spontaneously.
Also, on the ideas by R. Penrose and L. Di\'osi \cite{penrose1,
penrose2, penrose3, diosi0, diosi1, diosi2} who assumed that such
process should be connected to quantum aspects of gravitation.
There are other promising proposals based on Bohemian versions of
quantum theory applied to the inflationary field \cite{Bhomian},
but they will not be considered further in this work.

The simplest way this kind of process can be described is, by
assuming that at a certain stage in the cosmic evolution, there
was a self-induced jump in the state describing a particular mode
of the quantum field, in a manner that is similar to the quantum
mechanical reduction of the wave function associated with a
measurement. However, the reduction here is assumed to be
spontaneous and no external measuring device or observer is called
upon as triggering such collapse. A collapse scheme is a recipe to
characterize and select the state into which each of the modes of
the scalar field jumps at the corresponding time of collapse. The
collapse itself is described in a purely phenomenological manner,
without reference to any particular mechanism. As reported in, for
instance, \cite{pss, gabriel, landau}, the different collapse
schemes  generally give rise to different characteristic
departures from the conventional Harrison-Zel'dovich flat
primordial spectrum. There are, of course, more sophisticated
theories describing the collapse dynamics, such as those in
\cite{Bhom-Bub, Pearle:76, Pearle:79, PCS:13, GRW:85, GRW:86,
Pearle:89, GRW:90, jm, das}, however we will not consider those in the
present study, which is meant a first exploration of such ideas in
the context of the model proposed by Hollands and Wald.

The main objective of this article is to obtain the effects on the
shape of the primordial spectrum, that arise from a particular
collapse scheme, into the framework of the proposal of
\cite{wald}. We will show that with a simple collapse scheme and
for a certain range of values of the model parameters, one can
effectively recover a scale free spectrum. We will also consider
the dominant deviations from the flat spectrum that would arise in
this model.

The paper is organized as follows. In Section II, we develop the
necessary formalism of the model, describe the collapse scheme
implemented and show the results. Finally, in Section III, we make
our conclusions.

Throughout the work, we will use $c=1=\hbar$ and $l_p^2=\frac{8\pi
G}{3}$. Moreover, $l_0$ is a free parameter and its value is set at
the end of our calculations.

\section{The original model}

\subsection{The Einstein-fluid system}

Following the original work \cite{wald}, we start assuming that,
on spatial scales grater than $l_{0}$, the early universe is
dominated by a fluid with pressure $p$ and energy density $\rho$,
which are related by the equation of state $p=w\:\rho$ where $w\in
(0,1)$ is a constant. We will also assume that the background is
well described by a flat FRW metric.

Fixing the gauge (longitudinal) and restricting consideration to
the scalar modes, the perturbed metric in this case can be written
as
\begin{equation}\label{pert-metric}
ds^2 = a(\eta)^2 [-(1+2\Phi) d\eta^2 + (1-
2\Phi)\delta_{ij}dx^idx^j],
\end{equation}
where $\Phi=\Phi(\eta, x^{i})$ characterizes all deviations from
homogeneity and isotropy in the space-time.

The background is completely described by the value of $w$. In
particular, the scale factor is $a(\eta) = B
\eta^{\frac{2}{3w+1}}$, where $B$ is fixed so that $a(\eta_{{\rm
today}})=1$, and the background fluid density is $\rho \propto
a^{-3(w + 1)}$.

The next step in the analysis of this model, consists in the
quantization of the perturbations of the coupled Einstein-fluid
system. We will adhere to the treatment of the problem in
\cite{wald}, which follows the general approach formulated in
\cite{muk}. In that work, the authors focus on the variable $v$,
which is a linear combination of the fluid velocity potential
$\varphi_{v}$ and the metric potential $\Phi$,
\begin{equation}\label{variable}
v\equiv  \frac{1}{ \sqrt{6}\:l_p } ( \varphi_{v} -2\:z\:\Phi )
\end{equation}
where $z \equiv - \frac{a \sqrt{\beta} }{ \mathcal{H} c_{s} }$,
with $\beta \equiv \mathcal{H}^{2} - \mathcal{H'}$, the conformal
Hubble parameter is $\mathcal{H} \equiv \frac{a'}{a}$, and
$c_{s}=\sqrt{w}$ is the speed of sound in the fluid. Here, a prime
denotes partial derivative with respect to conformal time $\eta$.

The action up to second order in perturbation variables (Eq. (10.62) of \cite{muk}) reads,
\begin{equation}\label{accion}
\delta^{2} S = \frac{1}{2}\int d\eta\: d^3x \bigg( v'^2 -
c^{2}_{s}(\nabla v)^2 + \frac{z''}{z} v^2 \bigg)
\end{equation}
The above action is identical to the standard action of a free
scalar field ($v$) with a time-dependent mass. Moreover, $v$ is
analogous to the Mukhanov-Sasaki variable \cite{mukv}, and it is
the field that will be treated quantum mechanically.

Then, the Lagrangian density is
\begin{equation}\label{Lagrange}
\delta^{2} \mathcal{L} =  \frac{1}{2}\bigg( v'^2 -
c^{2}_{s}(\nabla v)^2 + \frac{z''}{z}v^2 \bigg),
\end{equation}
the momentum canonical to $v$ is $\pi \equiv
\partial \delta^{2}\mathcal{L}/\partial v'= v'$, and therefore the
Hamiltonian can be written as
\begin{equation}\label{Hamiltonian}
H(\eta)=\frac{1}{2}\int d{\bf x}\Big[\pi^{2} + c^{2}_{s}({\bf
\nabla}v)^{2}-\frac{z''}{z} v^2 \Big].
\end{equation}

In the quantization process, the field $v$ and its conjugate
momentum $\pi$ are promoted to operators acting on a Hilbert space
$\mathscr{H}$. These satisfy the standard equal time commutation
relations
\begin{eqnarray}\label{conmut}
\nonumber&& [\hat{v}(\eta,{\bf x}),\hat{\pi}(\eta,{\bf
x}')]=i\:\delta({\bf x}-{\bf x}') \\&& [\hat{\pi}(\eta,{\bf
x}),\hat{\pi}(\eta,{\bf x}')]=0=[\hat{v}(\eta,{\bf
x}),\hat{v}(\eta,{\bf x}')]
\end{eqnarray}

In the Heisenberg picture, the equations of motion are obtained
from
\begin{equation}\label{ecmov1}
i\hat{v}'=[\hat{v},\hat{H}],\:\:\:i\hat{\pi}'=[\hat{\pi},\hat{H}]
\end{equation}
that for the case of $\hat{v}$ it results to be
\begin{equation}\label{ecmov}
\hat{v}''+\Big(c_s^2\nabla^2-\frac{z''}{z}\Big)\hat{v}=0.
\end{equation}

The standard procedure is to write the general solution to this
equation decomposing $\hat{v}$ in terms of the time-independent
creation and annihilation operators. For practical reasons, we
will work with periodic boundary conditions over a box of size
$L$, where $k_i L=2\pi n_i$ for $i=1,2,3$. So we write
\begin{equation}\label{descomp}
\hat{v}(\eta,{\bf
x})=\frac{1}{\sqrt{2}\:L^{3/2}}\:\sum_{k\ne0}\hat{v}_{\bf k}(\eta)
\:e^{i{\bf{k}}\cdot {\bf{x}}}
\end{equation}
with $\hat{v}_{\bf k}(\eta)=\hat{a}_{\bf
k}\:v_k(\eta)+\hat{a}_{-\bf k}^\dagger\:v_k^\ast(\eta)$, and the
normal modes $v_k(\eta)$ satisfying
\begin{equation}\label{ecmovmodos}
v_k''+\Big(k^2 c_s^2-\frac{z''}{z}\Big)v_k=0.
\end{equation}
The normalization for the modes $v_k(\eta)$ is chosen such that
\begin{equation}\label{normaliz}
v_k(\eta)\: v'^{\ast}_{k}(\eta)- v'_{k}(\eta)\:
v_k^{\ast}(\eta)=2i,
\end{equation}
which leads to the standard commutation relations for the creation
and annihilation operators,
\begin{eqnarray}\label{conmut2}
\nonumber&& [\hat{a}_{\bf k},\hat{a}^\dagger_{\bf
k'}]=\delta_{{\bf k}{\bf k'}}
\\&& [\hat{a}_{\bf k},\hat{a}_{\bf k'}]=0=[\hat{a}^\dagger_{\bf k},\hat{a}^\dagger_{\bf k'}]
\end{eqnarray}
and the Fock space can be constructed in the standard way starting
with the vacuum state, i.e. the state defined by $\hat{a}_{\bf
k}|0\rangle=0$ for all ${\bf k}$. The choice of the functions
$v_k(\eta)$ corresponds to the selection of the vacuum  state for
the quantum field.

Following \cite{wald}, we choose as initial state, the vacuum
whose mode functions $v_k(\eta)$ satisfy (\ref{ecmovmodos}) at the
birth time $\eta_{0}^{k}$ (i.e. the time of birth of different
modes are distinct), and are given by
\begin{equation}\label{vacio}
v_k(\eta^{k}_{0})=\frac{1}{\sqrt{k c_s}}e^{-ikc_s\eta^{k}_{0}}\:\:
{\rm{and}} \:\:\: v'_k(\eta^{k}_{0})=-i \sqrt{k
c_s}e^{-ikc_s\eta^{k}_{0}}
\end{equation}
That is, the modes are set in the instantaneous vacuum state
corresponding at the specific time of birth of each mode.

Using that $\frac{z''}{z}=\frac{a''}{a}$ and defining
$A=\frac{2(1-3w)}{(3w+1)^2}$, the equation (\ref{ecmovmodos}) can
be re-written as,
\begin{equation}\label{ecmovmodos2}
v_k''+\Big(k^2 c_s^2-\frac{A}{\eta^2}\Big)v_k=0
\end{equation}
whose general solution is
\begin{equation}\label{modos}
v_k(\eta)=\sqrt{\eta}\Big[C_1 J_n(k c_s \eta)+C_2 Y_n(k c_s
\eta)\Big]
\end{equation}
In the last equation, $J_n$ and $Y_n$ are the Bessel functions of
the first and second kind respectively, and
$n=\frac{1}{2}\sqrt{1+4A}$. On the other hand, the constants $C_1$
and $C_2$ are obtained from the initial data at time $\eta_{0}^k$
according to (\ref{vacio}). Evaluating these constants we find
\begin{eqnarray}\label {c1c2}
C_{1}&=& \frac{\pi}{4\sqrt{c_{s}k\:\eta^{k}_{0}}}\Big[(2n + 1 +
2\:ic_{s}k\:\eta^{k}_{0})\:Y_{n}(c_{s}k\:\eta^{k}_{0})+ \nonumber\\
 &-& 2\:c_{s}k\:\eta^{k}_{0}\:Y_{n+ 1}(c_{s}k\:\eta^{k}_{0})
\Big]e^{-ic_{s}k\eta^{k}_{0}}\nonumber\\
C_{2}&=& -\frac{\pi}{4\sqrt{c_{s}k\:\eta^{k}_{0}}}\Big[(2n + 1 +
2\:ic_{s}k\:\eta^{k}_{0})J_{n}(c_{s}k\:\eta^{k}_{0})+ \nonumber\\
 &-&2\:c_{s}k\:\eta^{k}_{0}\:J_{n+1}(c_{s}k\:\eta^{k}_{0})
\Big]e^{-ic_{s}k\eta^{k}_{0}}.
\end{eqnarray}

In order to connect with observations, we must relate our quantum
variable, characterizing the perturbation of the coupled
Einstein-fluid system through the field $v$, with the
gravitational potential $\Phi$. This is done using the equation
(Eq. (12.8) of \cite{muk})
\begin{equation}\label{EinsteinP}
\nabla^{2} \Phi = -\sqrt{\frac{3}{2}}\frac{l_{p}\beta}{\mathcal{H}
c_{s}^{2}}\bigg(\frac{v}{z} \bigg)'
\end{equation}

The final hypothesis of the model concerns the time of birth of
the modes (as mentioned above). The mode $k$ is born when the
scale factor is such that satisfies
\begin{equation}\label{Birth}
 a_0 =k\:l_0.
\end{equation}

Then one considers, in the standard manner, the spectrum
characterized by the two point function of the field $\hat{v}$ in
the vacuum state $|0\rangle$ associated with the above specified
mode functions. The point is that, for modes with wavelength
greater than the Hubble radius, at the time of decoupling, one
finds as in \cite{wald} a scale free spectrum of density
perturbations:
\begin{equation}\label{espwald}
\mathcal{P}_{\Phi}(k)\sim\frac{l_p^2}{l_0^2}\frac{3w^{1/2}(6w+5)}{4(3w+5)^2}\frac{1}{k^3}
\end{equation}

However, as was mentioned in the Introduction, it is easy to
see that this state $|0\rangle$, is perfectly homogeneous and
isotropic (i.e. it is annihilated by the quantum generators of
rotations and translations). As the dynamical evolution preserves
such symmetries, the state of the system will be symmetric
(homogeneous and isotropic) at all times. In fact, there is
nothing, given the standard unitary evolution of the quantum
theory, that could be invoked to avoid such conclusion. The issue
is then, how do we account for a universe with seeds of cosmic
structure, not to say an appropriate spectrum for such
inhomogeneities, starting from an isotropic and homogeneous state?
We will not discuss these conceptual issues in detail here, and we
direct the interested reader to see some relevant works in the
literature (e.g. \cite{pss, sud, conceptual}).

In the following sections, we will study the incorporation of the
collapse hypothesis, a proposal designed to address the conceptual
difficulty mentioned above, into this model. We will consider the
modifications on the predicted form of the spectrum that results
from the analysis that incorporates such a collapse. As we will
see, under certain conditions, one can recover an almost scale
free spectrum for scalar perturbations, but generically there
would be some characteristic deviations thereof.

\subsection{Incorporating the collapse}

In the approach taken so far, both the metric and the fluid
perturbations are treated at the quantum level. Then, from
(\ref{EinsteinP}), we must also promote $\Phi$ to a quantum
operator. Our focus will be the scalar metric perturbation $\Phi$,
being this the link with the observations, representing the small
anisotropies in the temperature of the CMB. What we need is to
find an expression representing the relevant metric perturbation
$\Phi$.

We will assume that the classical description is only relevant for
those particular states for which the quantity in question is
sharply peaked, and that the classical description corresponds to
the expectation value of said quantity.

Following previous works (e.g. \cite{gabriel}), where the case has
been addressed in detail, we will assume the validity of the
identification
$\Phi^{\Xi}(x)\equiv\langle\Xi|\hat{\Phi}(x)|\Xi\rangle\equiv\langle\hat{\Phi}(x)\rangle_{\Xi}$,
with $|\Xi\rangle$ the corresponding state of the quantum field.
That in turn will be either the pre-collapse vacuum state or the
post-collapse state of the field $\hat{v}(x)$, characterizing
jointly the metric and fluid perturbation, and being sharply
peaked in the associated variable $\hat{\Phi}(x)$.

Calculating its expectation value in the vacuum state, in Fourier
space, we obtain
\begin{equation}\label{phiop}
\langle0|\hat{\Phi}_{\bf
k}(\eta)|0\rangle=\sqrt{\frac{3}{2}}\frac{l_{p}\:\beta}{\mathcal{H}
c_{s}^{2}}\frac{1}{k^2}\bigg(\frac{\langle0|\hat{v}_{\bf
k}(\eta)|0\rangle}{z} \bigg)'
\end{equation}

Therefore, given that $\langle0|\hat{v}_{\bf k}(\eta)|0\rangle=0$,
and without invoking something like a collapse, we will have that
$\langle0|\hat{\Phi}_{\bf k}(\eta)|0\rangle=0$ at all times. This
explicitly exhibits the above alluded problem, that the symmetries
of the initial state do not seem to be compatible with any
perturbation of the metric.

Just as in preceding works \cite{pss, sud}, we are lead to
introduce something like what we call a \emph{collapse
hypothesis}. When the modes are born at time $\eta_0^{k}$, the
state $|0\rangle$ is perfectly symmetric. At some collapse time,
$\eta^{k}_{c}$, a transition to a new state
$|0\rangle\rightarrow|\Xi\rangle$ is produced, which does not have
the initial symmetries. And in this new state, we will have that
$\langle\Xi|\hat{v}_{\bf k}(\eta)|\Xi\rangle\ne0$ for all
$\eta\ge\eta^{k}_{c}$, and therefore, the collapse process will
generate the seeds of cosmic structure.

That process is thought to represent some novel aspect of physics,
connecting with properties of quantum gravity, as has been
suggested, for instance, by Di\'osi and Penrose \cite{penrose1,
penrose2, penrose3, diosi0, diosi1, diosi2}. There is a long
history of studies about proposals involving something like a
collapse of the wave function. Most of them, from the community
working on foundations of quantum theory (see e.g. \cite{bassi}).
The manner in which this problematic issues appear in the
inflationary cosmological context was considered first in
\cite{pss}.

In this work, we will consider the simple case where only one
collapse happens per mode ${\bf k}$. Specifically, we will assume
that, at $\eta^{k}_{c}$, a given mode ${\bf k}$ undergoes a
collapse $|0_{\bf k}\rangle\rightarrow|\Xi_{\bf k}\rangle$. These
collapses will be assumed to take place according to certain
specific rules which we will describe in detail and, as we will
see, they will depend on a particular \emph{collapse scheme}
considered, and will induce a change in the expectation value of
$\hat{\Phi}_{\bf k}(\eta)$. The important point here is that,
after the collapse of the mode ${\bf k}$, the universe will be no
longer homogeneous and isotropic, in regards to that mode.

While the issues related to the transition from a quantum
description to a classical one have been the subject of debate
(e.g. \cite{sud, stat}), and the collapse proposal seems better
suited to the semiclassical treatments (where the matter fields
are quantized and the metric perturbations are not, e.g.
\cite{alberto}), here we will follow the most usual approach in
which the composite field $v$ is quantized (see, for instance
\cite{gabriel}).

Thus, our equation for the perturbation $\Phi$ and for
$\eta\ge\eta^{k}_{c}$ (in Fourier space) will be:
\begin{eqnarray}\label{uno}
\nonumber\Phi^{\Xi}_{\bf k}(\eta)&&\equiv\langle\hat{\Phi}_{\bf
k}(\eta)\rangle_{\Xi}\equiv\langle\Xi|\hat{\Phi}_{\bf
k}(\eta)|\Xi\rangle=\\&&=\sqrt{\frac{3}{2}}\frac{l_{p}\:\beta}{\mathcal{H}
c_{s}^{2}}\frac{1}{k^2}\bigg(\frac{\langle\hat{v}_{\bf
k}(\eta)\rangle_{\Xi}}{z} \bigg)'
\end{eqnarray}

This is the point where we must make contact with the
observations.

The small anisotropies observed in the temperature of the CMB
radiation, $\delta T(\theta,\phi)/T_0$, can be described in terms
the coefficients $a_{lm}$ of the multipolar expansion
\begin{eqnarray}\label{dos}
\nonumber\frac{\delta T}{T_0}(\theta,\phi)&=&\sum_{l,m}a_{lm}\:
Y_{lm}(\theta,\phi)\\a_{lm}&=&\int\frac{\delta
T}{T_0}(\theta,\phi)\:Y^{\ast}_{lm}(\theta,\phi)\: d\Omega
\end{eqnarray}
Here, $\theta$ and $\phi$ are the two-sphere coordinates,
$Y_{lm}(\theta,\phi)$ are the spherical harmonics, and $T_0=2.725$
K is the CMB average temperature today. At large angular scales
(low $l$) the Sachs-Wolfe effect is the dominant source to the CMB
anisotropies. It is well known that for this case,
\begin{equation}\label{swolfe}
\frac{\delta T}{T_0}(\theta,\phi)=\frac{1}{3}\Phi(\eta_D,R_D)
\end{equation}
where $\eta_D$ and $R_D$ are evaluated in the decoupling epoch.

Now, expanding to $\Phi(\eta_D,R_D)$ in Fourier modes and using
the spherical Bessel functions of order $l$, $j_l(k\: R_D)$, the
expression (\ref{dos}) can be written as:
\begin{equation}\label{alm}
a_{lm}=\frac{4\pi i^l}{3L^{3/2}}\sum_{\bf k}
j_l(kR_D)\:Y^{\ast}_{lm}(\hat{k})\Phi_{\bf k}^{\Xi}(\eta_D)
\end{equation}

The individual complex quantities $a_{lm}$, correspond to sums of
complex contributions $\Phi_{\bf k}^{\Xi}(\eta_D)$, each one
having a certain randomness, but leading in combination to a
characteristic value in just the same way as a two-dimensional
random walk made of multiple steps. As in any random walk, the
only thing that can be done is to calculate the most likely (ML)
value for the total displacement, with the expectation that the
observed quantity will be close to that value. Thus, one needs to
estimate the most likely value of the quantity
\begin{eqnarray}\label{alm2}
\nonumber&&|a_{lm}|^2=\\
\nonumber&&\frac{16\pi^2}{9L^{2}}\sum_{\bf k,k'}
j_l(kR_D)\:j_l(k'R_D)Y^{\ast}_{lm}(\hat{k})Y_{lm}(\hat{
k'})\Phi_{\bf k}^{\Xi}(\eta)\Phi_{\bf k'}^{\Xi\ast}(\eta)\:\:\:\:.
\end{eqnarray}

Following the approach first used in \cite{pss}, one accomplishes
this with the help of an imaginary ensemble of universes (each one
corresponding to a possible realization of the collapse). We will
identify the most likely value with the observed one, and will
estimate its value using the ensemble mean value of the
corresponding quantity. That is,
$|a_{lm}|^2_{\rm{ML}}\simeq|a_{lm}|^2_{\rm{obs}}\simeq\overline{|a_{lm}|^2}$.

Thus, one finds that the quantity that contains the information
regarding the theoretical prediction for the primordial spectrum,
and that, in an effective way, would correspond to the expression
(\ref{espwald}), will be obtained once we calculate, for the
relevant modes, the quantity $\overline{\Phi_{\bf
k}^{\Xi}(\eta)\Phi_{\bf k'}^{\Xi\ast}(\eta)}$, where the over-line
indicates the ensemble average. The relevant quantities for the
analysis of the seeds of cosmic structure are those characterizing
the statistics of the collapse, as we will see next. Finally, a
direct connection with the standard results would be obtained if
we write,
\begin{equation}\label{tres}
\overline{\Phi_{\bf k}^{\Xi}(\eta)\Phi_{\bf
k'}^{\Xi\ast}(\eta)}=\mathcal{P}_{\Phi}(k)\:\delta_{{\bf k}{\bf
k'}}
\end{equation}

Our aim now is to analyze the modifications in the predictions for
the primordial scalar fluctuation spectrum that result from
incorporating the collapse hypothesis into the model developed in
\cite{wald}.

We will consider weather, and under what circumstances, one can
obtain a prediction for a scale free spectrum for those
perturbations, for the case of the observationally relevant modes,
which have wavelengths greater than the Hubble radius at the time
of decoupling (and therefore, within this model, at all previous
times).

\subsection{Results}

In order to proceed to find our results, firstly we decompose the
operators $\hat{v}_{\bf k}(\eta)$ and $\hat{\pi}_{\bf k}(\eta)$ in
their real and imaginary parts,
\begin{eqnarray}\label{reim}
\nonumber\hat{v}_{\bf k}(\eta)&=&\hat{v}_{\bf
k}^{\rm{R}}(\eta)+i\:\hat{v}_{\bf k}^{\rm{I}}(\eta)\\
\nonumber\hat{\pi}_{\bf k}(\eta)&=&\hat{\pi}_{\bf
k}^{\rm{R}}(\eta)+i\:\hat{\pi}_{\bf k}^{\rm{I}}(\eta)
\end{eqnarray}
with $\hat{v}_{\bf
k}^{\rm{R,I}}(\eta)=\frac{1}{\sqrt{2}}\Big(\hat{a}_{\bf
k}^{\rm{R,I}}v_k(\eta)+\hat{a}_{\bf
k}^{\rm{R,I}\dagger}v_k^{\ast}(\eta)\Big)$ and $\hat{\pi}_{\bf
k}^{\rm{R,I}}(\eta)=\frac{1}{\sqrt{2}}\Big(\hat{a}_{\bf
k}^{\rm{R,I}}\pi_k(\eta)+\hat{a}_{\bf
k}^{\rm{R,I}\dagger}\pi_k^{\ast}(\eta)\Big)$, where $\hat{a}_{\bf
k}^{\rm{R}}=\frac{1}{\sqrt{2}}(\hat{a}_{\bf k}+\hat{a}_{\bf -k})$
and $\hat{a}_{\bf k}^{\rm{I}}=\frac{-i}{\sqrt{2}}(\hat{a}_{\bf
k}-\hat{a}_{\bf -k})$. Note that according to (\ref{accion}),
$\pi_k(\eta)=v'_k(\eta)$. In this manner, $\hat{v}_{\bf
k}^{\rm{R,I}}(\eta)$ and $\hat{\pi}_{\bf k}^{\rm{R,I}}(\eta)$ are
Hermitian operators. But now, the commutation relations will be:
\begin{eqnarray}\label{conmut3}
\nonumber [\hat{a}_{\bf k}^{\rm{R}},\hat{a}^{{\rm R}\dagger}_{\bf
k'}]&=&(\delta_{{\bf k},{\bf k'}}+\delta_{{\bf k},{\bf -k'}})
\\ \nonumber [\hat{a}_{\bf k}^{\rm{I}},\hat{a}^{{\rm I}\dagger}_{\bf
k'}]&=&(\delta_{{\bf k},{\bf k'}}-\delta_{{\bf k},{\bf -k'}})
\end{eqnarray}
with all the other commutators vanishing.

As was mentioned in the Introduction, we assume, in analogy
with standard quantum mechanics, that the collapse is somehow
analogous to an imprecise measurement (of the Hermitian operators
$\hat{v}_{\bf k}^{\rm{R,I}}$ and $\hat{\pi}_{\bf k}^{\rm{R,I}}$).

The proposal for the collapse scheme, assumes that at certain time
$\eta^{k}_{c}$, which can depend on the mode, the state
corresponding to the mode ${\bf k}$ undergoes an instantaneous
change (or collapse) so that the expectation values of the field
operators after such change become:
\begin{eqnarray}\label{cuatro}
\nonumber \langle\hat{v}_{\bf
k}^{\rm{R,I}}(\eta^{k}_{c})\rangle_{\Xi}&=&x_{\bf
k}^{\rm{R,I}}\sqrt{\langle(\Delta \hat{v}_{\bf
k}^{\rm{R,I}}(\eta^{k}_{c}))^2\rangle_0}\\
&=&\frac{x_{\bf
k}^{\rm{R,I}}}{\sqrt{2}}|v_k(\eta^{k}_{c})|\equiv s_{(k)}\\
\label{cuatrob}\langle\hat{\pi}_{\bf
k}^{\rm{R,I}}(\eta^{k}_{c})\rangle_{\Xi}&=& 0,
\end{eqnarray}
where $\langle(\Delta \hat{v}_{\bf
k}^{\rm{R,I}}(\eta^{k}_{c}))^2\rangle_0$ is the quantum
uncertainty of the operator $\hat{v}_{\bf k}^{\rm{R,I}}$ in the
vacuum state $|0\rangle$ at time $\eta^{k}_{c}$. This is one of the simplest
collapse schemes one can consider, where the collapse process changes the expectation
value of the field operators but leaves the expectation values of the momenta unchanged.
Other schemes are of course possible, such as when the collapse operators affect the later and not the former, or when the collapse process modifies both simultaneously, either preserving, or not, certain correlations present in the quantum state. Previous analysis considering the various kinds of collapse in standard inflationary models indicate that different alternatives lead to similar (but not exactly identical) results. The interested reader can see these differences, in e.g. \cite{adolfo, landau}. We will focus our attention here on the simple scheme specified above, that might be viewed as the most straightforward generalization of the collapse into the configuration variable (that is the position operators) and does not modify the expectation value of the conjugate momentum operators, which are used in the context of non-relativistic quantum mechanics \cite{Bhom-Bub, Pearle:76, Pearle:79, PCS:13, GRW:85, GRW:86, Pearle:89, GRW:90}.

The numbers $x_{\bf k}^{\rm{R,I}}$ are a collection of independent random
quantities (selected from a Gaussian distribution centered at zero
with unit-spread) that, as we will see next, will help determine
$\Phi_{\bf k}^{\Xi}(\eta)$ as a kind of random walk. This collapse
scheme is a particular case of other more general ones (see for
instance, \cite{gabriel}). As previously mentioned, we will assume
that we can estimate the most likely value of the random walk
displacement by its ensemble average that is
$|a_{lm}|^2_{\rm{ML}}\simeq\overline{|a_{lm}|^2}$. Therefore, in
this case, we can use
\begin{eqnarray}\label{xs1}
\overline{x_{\bf k}^{\rm{R}}x_{\bf k'}^{\rm{R}}}&=&\delta_{{\bf
k},{\bf k'}}+\delta_{{\bf k},{\bf
-k'}}\\\label{xs2}\overline{x_{\bf k}^{\rm{I}}x_{\bf
k'}^{\rm{I}}}&=&\delta_{{\bf k},{\bf k'}}-\delta_{{\bf k},{\bf
-k'}}.
\end{eqnarray}
We must emphasize that our universe corresponds to a single
realization of each of these random variables and, therefore, the
quantities of interest have a single specific value.

In order to compute $\Phi_{\bf k}^{\Xi}(\eta)$, we need to know
the quantity $\langle \hat{v}_{\bf
k}^{\rm{R,I}}(\eta)\rangle_{\Xi}$ for all $\eta\ge\eta^{k}_{c}$.
In order to obtain this, we make use of Ehrenfest equations,
\begin{eqnarray}\label{cinco}
\frac{d}{d\eta}\langle\hat{v}_{\bf
k}^{\rm{R,I}}(\eta)\rangle_{\Xi}&=&\langle\hat{\pi}_{\bf
k}^{\rm{R,I}}(\eta)\rangle_{\Xi}\\
\label{seis}\frac{d}{d\eta}\langle\hat{\pi}_{\bf
k}^{\rm{R,I}}(\eta)\rangle_{\Xi}&=&-\Big(c_s^2k^2-\frac{A}{\eta^2}\Big)\:\langle\hat{v}_{\bf
k}^{\rm{R,I}}(\eta)\rangle_{\Xi}
\end{eqnarray}
Differentiating (\ref{cinco}) with respect to $\eta$, and using
(\ref{seis}), we can write,
\begin{equation}\label{cincobis}
\frac{d^2}{d\eta^2}\langle\hat{v}_{\bf
k}^{\rm{R,I}}(\eta)\rangle_{\Xi}=-\Big(c_s^2k^2-\frac{A}{\eta^2}\Big)\:\langle\hat{v}_{\bf
k}^{\rm{R,I}}(\eta)\rangle_{\Xi}
\end{equation}
whose general solution is
\begin{equation}\label{siete}
\langle\hat{v}_{\bf
k}^{\rm{R,I}}(\eta)\rangle_{\Xi}=\alpha_1^{\rm{R,I}}f(\eta)+\alpha_2^{\rm{R,I}}h(\eta)
\end{equation}
hence,
\begin{equation}\label{ocho}
\langle\hat{\pi}_{\bf
k}^{\rm{R,I}}(\eta)\rangle_{\Xi}=\alpha_1^{\rm{R,I}}f'(\eta)+\alpha_2^{\rm{R,I}}h'(\eta).
\end{equation}

Here we have defined:
\begin{eqnarray}\label{nuevea}
f(\eta)&\equiv&\sqrt{\eta}\:J_n(c_sk\eta)\\
\label{nueveb}h(\eta)&\equiv&\sqrt{\eta}\:Y_n(c_sk\eta)
\end{eqnarray}

Evaluating (\ref{siete}) and (\ref{ocho}) at $\eta^{k}_{c}$, and
using our collapse scheme (\ref{cuatro}-\ref{cuatrob}), the
constants $\alpha_1^{\rm{R,I}}$ and $\alpha_2^{\rm{R,I}}$
(depending on $\eta^{k}_{c}$) are found to be,
\begin{eqnarray}\label{diez}
\alpha_1^{\rm{R,I}}&=&\frac{\pi s_{(k)}h'(\eta_c^{k})}{2}\\
\label{once}\alpha_2^{\rm{R,I}}&=&-\frac{\pi
s_{(k)}f'(\eta_c^{k})}{2}
\end{eqnarray}
and therefore:
\begin{equation}\label{doce}
\langle\hat{v}_{\bf k}^{\rm{R,I}}(\eta)\rangle_{\Xi}=\frac{\pi
s_{(k)}}{2}\Big[h'(\eta^{k}_{c}) f(\eta)-f'(\eta^{k}_{c})
h(\eta)\Big]
\end{equation}

We restrict our analysis to the cases where the time of birth, the
time of collapse for each mode, and the overall evaluation time
(the time of interest $\eta$, which strictly speaking should be
the decoupling time), satisfy that $\eta_0^{k} < \eta^{k}_{c} <
\eta$. Furthermore, we recall that we are interested in the
observationally relevant modes having $k\eta \ll 1$ corresponding
to wavelengths greater than the Hubble radius at the time of
decoupling.

Thus, we can use the asymptotic forms for small arguments of
Bessel functions, both in the functions $h'(\eta^{k}_{c})$ and
$f'(\eta^{k}_{c})$, and for $f(\eta)$ and $h(\eta)$ at the end of
the calculation.

In order to proceed, we make some numerical estimates. By taking
$w=0.3$ and $k\sim10^{-3}\:\rm{Mpc^{-1}}$ we find that
$\eta_0^{k}\sim10^{-50}\:\rm{Mpc}$. Therefore, (\ref{doce}) can be
approximated by,
\begin{eqnarray}\label{trece}
\nonumber\langle\hat{v}_{\bf
k}^{\rm{R,I}}(\eta)\rangle_{\Xi}&\simeq&\frac{\pi
s_{(k)}}{2}\bigg[\frac{\sqrt{2}\Gamma(n)(n-\frac{1}{2})}{\pi\sqrt{k
c_s}\eta^{k}_{c}}\Big(\frac{2}{k
c_s\eta^{k}_{c}}\Big)^{n-\frac{1}{2}}
f(\eta)+\\
&-&\frac{\sqrt{k c_s}
(n+\frac{1}{2})}{\sqrt{2}\Gamma(n+1)}\Big(\frac{k
c_s\eta^{k}_{c}}{2}\Big)^{n-\frac{1}{2}} h(\eta)\bigg]
\end{eqnarray}

We have performed separate numerical analysis of this issue and
have found that the exact results are essentially
indistinguishable from the ones obtained with the above
approximations.

Next, we note that for the relevant modes, one can approximate
$|v_k(\eta)|$ from (\ref{modos}) by the expression
\begin{equation}\label{modoaprox}
|v_k(\eta)|\simeq \frac{1}{4n\sqrt{c_{s}k}}\Big[ (2n-1)\Big(
\frac{\eta}{ \eta_{0}^k } \Big)^{n+\frac{1}{2}} + (1+2n)\Big(
\frac{\eta_{0}^k}{ \eta } \Big)^{n -\frac{1}{2}}  \Big],
\end{equation}
which, in combination with the hypothesis of \cite{wald}, i.e. the
modes are born when $a(\eta_0^{k})\simeq k l_0$, allow us to write
\begin{equation}\label{modoaprox2}
|v_k(\eta)|\simeq\frac{a(\eta)}{4n  l_0 \sqrt{c_{s}} k^{3/2}
}\Big[(2n - 1) + (2n +1) \Big( \frac{kl_0}{a(\eta)} \Big)^{
\frac{4n}{2n+1} } \Big],
\end{equation}
where we can see that if $\eta = \eta_{0}^k$, then
$|v_k(\eta_{0}^k)| = \frac{1}{\sqrt{c_{s}k}}$. While that for the
case of pure radiation, this means $w=\frac{1}{3}$ (or
$n=\frac{1}{2}$), $|v_k(\eta)|$ becomes $ |v_k(\eta)| =
\frac{1}{\sqrt{c_{s}k}}$ for all $\eta$.

Next, using (\ref{cuatro}), we evaluate the quantum uncertainty of
$\hat{v}_{\bf k}^{\rm{R,I}}(\eta^{k}_{c})$ in the state
$|0\rangle$. On the other hand, an approximation for
$\langle(\Delta \hat{v}_{\bf k}^{\rm{R,I}}(\eta))^2\rangle_0$,
evaluated at $\eta^{k}_{c}$ will be,
\begin{eqnarray}\label{catorce}
\langle(\Delta \hat{v}_{\bf
k}^{\rm{R,I}}(\eta_c^{k}))^2\rangle_0&\simeq&\frac{a^{2}(\eta^{k}_{c})}{32n^{2}l^{2}_{0}c_{s}k^{3} }\Big[(2n - 1)+ \nonumber\\
&+&(1+2n) \Big(\frac{kl_{0}}{a(\eta^{k}_{c})}\Big)^{\frac{4n}{2n
+1}}  \Big]^2.
\end{eqnarray}

Now, we note that from the above expression, together with
(\ref{cuatro}), one obtains
\begin{eqnarray}\label{Skaprox2}
s_{(k)}&\simeq&\frac{a(\eta^{k}_{c})}{4n\sqrt{2c_{s}}l_{0} k^{3/2} }\Big[2n - 1 +\nonumber\\
&+& (1+2n) \Big(\frac{kl_{0}}{a(\eta^{k}_{c})}\Big)^{\frac{4n}{2n
+1}} \Big]\:x_{\bf k}^{\rm{R,I}}.
\end{eqnarray}

Given that $z=\gamma\: \eta^{\frac{2}{3w+1}}$, with
$\gamma=\frac{-\sqrt{3w+3}}{\sqrt{2}\:c_s}\:\eta_{{\rm
today}}^{\frac{-2}{3w+1}}$, the expression (\ref{uno}) can now be
rewritten as
\begin{equation}\label{dieciseis}
\langle\hat{\Phi}_{\bf
k}^{\rm{R,I}}(\eta)\rangle_{\Xi}=\sqrt{\frac{3}{2}}\frac{l_{p}\:\beta}{\gamma\:\mathcal{H}
c_{s}^{2}}\frac{1}{k^2}\bigg(\frac{\langle\hat{v}_{\bf
k}^{\rm{R,I}}(\eta)\rangle_{\Xi}}{\eta^{\frac{2}{3w+1}}} \bigg)'
\end{equation}

Then, by using (\ref{trece}) along with (\ref{Skaprox2}) into
(\ref{dieciseis}), and using the fact that the random variables
$x_{\bf k}^{\rm{R,I}}$ satisfy (\ref{xs1}-\ref{xs2}), we can
finally write,
\begin{eqnarray}\label{diezsiete}
\nonumber&&\overline{\Phi_{\bf k}^{\Xi}(\eta)\Phi_{\bf
k'}^{\Xi\ast}(\eta)}=2\:\overline{\langle\hat{\Phi}_{\bf
k}^{\rm{R,I}}(\eta)\rangle_{\Xi}\:\langle\hat{\Phi}_{\bf
k'}^{\rm{R,I}}(\eta)\rangle_{\Xi}}\simeq\\
&&\simeq\frac{3\:l_{p}^{2}\:\beta^{2}a^{2}(\eta_{c}^k)}{256\:n^{2}
\mathcal{H}^{2} c_{s}^{5}\:l_{0}^{2}\:\gamma^{2} k^3}\Big[2n - 1 +\nonumber\\
&+&  (1+2n) \Big(\frac{kl_{0}}{a(\eta^{k}_{c})}\Big)^{\frac{4n}{2n
+1}} \Big]^2 \Bigg[ \delta_{1} + \frac{\delta_{2}}{k^{2}}
  \Bigg]^2\:\delta_{{\bf k}{\bf
k'}}\:\:\:\:\:\:\:\:
\end{eqnarray}
where:
\begin{eqnarray}\label{deltas}
\delta_{1}&\equiv&\Big(n - \frac{1}{2}\Big)\frac{c_{s}^{2}\: \eta
}{n(n + 1)2^{n+ 1}}\Big(\frac{2}{\eta^{k}_{c}}\Big)^{n +
\frac{1}{2}}\nonumber\\
\delta_{2}&\equiv&\Big(n +
\frac{1}{2}\Big)\frac{2^{n+1}}{\eta^{2n+
1}}\Big(\frac{\eta^{k}_{c}}{2}\Big)^{n - \frac{1}{2}}
\end{eqnarray}

Thus recalling that $a(\eta^{k}_{c})= B (\eta^{k}_{c})^{n+1/2}$ it
is easy to see that the right hand side of (\ref{diezsiete}),
takes the following form:
\begin{eqnarray}\label{simplificado}
\overline{\Phi_{\bf k}^{\Xi}(\eta)\Phi_{\bf k'}^{\Xi\ast}(\eta)}
&\simeq&\frac{3}{2048 k^{3} }\Bigg\{ \frac{ Bl_{p} \beta \eta  }{
( n + 1 )n^{2}  \sqrt{c_{s}}
\mathcal{H} l_{0} \gamma}  \Bigg[2n - 1+ \nonumber\\
&+& (1+2n) \Big(\frac{kl_{0}}{a(\eta^{k}_{c})}\Big)^{\frac{4n}{2n +1}}\Bigg]\Bigg[2n-1+ \nonumber\\
&+& \frac{4n(2n^{2}+ 3n+ 1) (\eta^{k}_{c})^{2n}}{c_{s}^{2}\eta^{2n
+ 2}k^{2}}\Bigg] \Bigg\}^{2}\delta_{{\bf k}{\bf k'}}
\end{eqnarray}

We note that, when $|2n -
1|\gg(1+2n)\Big(\frac{kl_{0}}{a(\eta^{k}_{c})}\Big)^{\frac{4n}{2n
+1}}$, and $|2n - 1|\gg\frac{4n(2n^{2}+
3n+1)(\eta^{k}_{c})^{2n}}{c_{s}^{2}k^{2}\eta^{2n+2}}$, the
dominant contribution would lead to the standard scale invariant
prediction for the spectrum $\overline{\Phi_{\bf
k}^{\Xi}(\eta)\Phi_{\bf k}^{\Xi\ast}(\eta)} \propto  \frac{1}{
k^{3} }$, and its value would be independent of the value of the
time of collapse $\eta^{k}_{c}$ of the mode ${\bf k}$. However, it
is clear that this will not be the case when $n=1/2$ corresponding
to the pure radiation case where $w=1/3$.

Next, we consider the corrections to the scale invariant
component. For this, we must focus on the two above indicated
inequalities. When $n \in (0,\frac{2}{3})$, we have $|2n -
1|<(1+2n)$. Then, in order to satisfy the first inequality, it is
necessary that
$\Big(\frac{kl_{0}}{a(\eta^{k}_{c})}\Big)^{\frac{4n}{2n +1}} \ll
1$. However, the exponent is such that $0<\frac{4n}{2n
+1}<\frac{3}{2}$. So, although $k l_0=
a(\eta^{k}_{0})<a(\eta^{k}_{c})$ is assured (modes would not
collapse before they are born), we need that the collapse to take
place sufficiently late so that the two scales differ by, say, at
least, one order of magnitude.

In short, we can conclude that the first inequality will be
satisfied as long as $0 \ll n < \frac{3}{2}$, and the collapse is
such that $a(\eta^{k}_{0})$ and $a(\eta^{k}_{c})$ differ by at
least one order of magnitude.

Let us focus now on the second inequality. The first issue we must
be concerned with, is the fact that $c_{s}$ goes to zero when $w$
goes to zero, or equivalently, when $n$ goes to $\frac{3}{2}$. The
term $(\eta^{k}_{c} /\eta )^{2n}$ is small simply because we are
interested on the quantity in (\ref{simplificado}) at very late
times; which clearly must be long after the collapse has taken
place for all the relevant modes. This term is further modulated
by a $4n(2n^{2}+ 3n+1)$ which lies in the range $(0,60)$. Thus,
the second inequality will be satisfied if $0 \ll c_{s} <1$ (this
means that $0 < n\ll\frac{3}{2}$), together with the conditions
that ensure that $ c_{s} k \eta$ is not too small. This, of
course, will be most easily satisfied the earlier the collapse for
the relevant modes occurs.

Therefore, in the case where the fluid is such that $0 \ll
n\ll\frac{3}{2}$ and thus $|2n - 1|\gg0$ (i.e. $0 \ll w \ll 1$ and
$|3w - 1|\gg0$)\footnote{Our use of the notation  '$\ll$' in this
discussion is meant to indicate that the inequality is satisfied
by a sufficiently large margin, to ensure the dominance of the
term in question by a large enough factor, given the values of the
other quantities appearing in the corresponding expressions, and
the precision of the existing observations.}, and where the
collapse occurs early enough, we find that the expression
(\ref{simplificado}) for the power spectrum is well approximated
by
\begin{eqnarray}\label{correciones}
\overline{\Phi_{\bf k}^{\Xi}(\eta)\Phi_{\bf k'}^{\Xi\ast}(\eta)}
&\simeq&\frac{3}{2048 k^{3} }\Bigg\{ \frac{ Bl_{p} \beta \eta (2n
- 1)^{2} }{ ( n + 1 )n^{2} \sqrt{c_{s}}
\mathcal{H} l_{0} \gamma}  \Bigg[ 1 + \nonumber\\
&+&  \frac{(1+2n)}{(2n - 1)}\Big(\frac{kl_{0}}{a(\eta^{k}_{c})}\Big)^{\frac{4n}{2n +1}}+\nonumber\\
&+& \frac{4n(2n^{2}+ 3n+ 1) (\eta^{k}_{c})^{2n}}{c_{s}^{2}\eta^{2n
+ 2}(2n - 1)k^{2}}  \Bigg] \Bigg\}^{2}\delta_{{\bf k}{\bf k'}}
\end{eqnarray}
which has a leading term corresponding to the standard flat
spectrum together with corrections associated with the collapse,
and which clearly depends on the collapse times for the relevant
modes.

This is very similar to what was found to occur in the standard
inflationary model with an instantaneous collapse. We can see this
by considering Eq. (88) of Ref. \cite{pss}:
\begin{equation}\label{C(k)dePSS}
C(k) = 1 + \frac{ 2 }{ (k\eta^{k}_{c})^{2} }\sin^{2}(\Delta_{k}) +
\frac{ 1 }{ k\eta^{k}_{c} }\sin(2\Delta_{k}),
\end{equation}
with $\Delta_{k} = k\eta_{D} - k\eta^{k}_{c}$, where $\eta_{D}$ is
the conformal time at decoupling, and $C(k)$ is a dimensionless
function which encodes all the information about the effects of
the details of the collapse scheme on the observational power
spectrum. As discussed in that work, it is only if $C(k)$ is
independent of $k$ that the standard scale invariant spectrum is
recovered. Now, considering only the modes of interest, and taking
into account that in the inflationary situation
$k\eta_{D}\approx0$, then as long as $0<|k\eta^{k}_{c}|\ll1$
(which would be the case if the collapse of the relevant modes
takes place near the end of the inflationary period), the quantity
$\Delta_{k}\rightarrow -k\eta^{k}_{c}$, and therefore performing a
series expansion in (\ref{C(k)dePSS}) one finds that,
\begin{equation}\label{answtaylor}
C(k) = 1 + \Big[\frac{2(k\eta^{k}_{c})^{2}}{3} + O(3)\Big].
\end{equation}
Then, as long as for the relevant modes $|k\eta^{k}_{c}|\ll1$, the
dominate terms would lead to a standard scale free spectrum. In
that case, just as it happens in the situation considered in this
work, (see (\ref{diezsiete})), the dominant part of
(\ref{answtaylor}) is independent of the value of the collapse
time $\eta^{k}_{c}$ of the mode ${\bf k}$.

At this point, it is worth mentioning that once the wave function has collapsed, it keeps evolving according
to the Schr\"{o}dinger equation and thus it will not remain very well localized even if it was so right after the collapse.
The general issue of how well localized should be the field and its momentum conjugate at various epochs is one that needs a careful study, which should involve, among other things, the possibility of multiple collapses of each mode. In this regard, we should mention that the analysis contemplating a multiplicity of collapses, within the more traditional inflationary paradigm, has been carried out in \cite{multiples} indicating that avoiding important deviations from the simple single collapse scheme require some limitation in the number of collapses per mode.

However, one thing which has to be mentioned is that the sharpness in the value of the directly observable quantities (e.g. $a_{lm}$), is likely to be much bigger than what might be inferred directly from the spread of the individual modes of $v_{\bf k}$. The reason for this is that, as indicated in equation (\ref{alm}), the quantity of interest is the sum of a very large number of terms (one for each wavenumber ${\bf k}$), and thus, one should expect something like the standard $1/\sqrt{N}$ decrease in the variance that is characteristic of the statistics of aggregates of large number, $N$, of uncorrelated contributions. The situation is of course more complicated, among other reasons, due to the different weights with which the various random terms appear in the sum (\ref{alm}).  We will leave for future works the detail analysis of this and related issues.

In the case $w\approx \frac{1}{3}$, the model does not lead to the
standard scale free spectrum and the details of the predicted
shape will depend strongly on the model parameters and the
assumptions made regarding the collapse times $\eta^{k}_{c}$ for
the observationally relevant modes with
$k\in[10^{-4},10^{-1}]\:{\rm{Mpc^{-1}}}$.

From the original proposal \cite{wald}, it can be inferred that
for the case of pure radiation ($w=1/3$), the scale free spectrum
cannot be recovered. We are naturally lead to consider whether the
introduction of the collapse hypothesis could modify that
conclusion. This possibility arises because, as we saw, the
collapse brings an extra set of parameters represented by the
collapse times of each of the modes. This can be explored by
setting $w=1/3$, ($n=1/2$) in equation (\ref{simplificado}). We
find that,
\begin{equation}\label{radiac}
\mathcal{P}_{\Phi}^{w=1/3}(k)\simeq\frac{3 \:l_p^2\: \beta^2}{2
\mathcal{H}^{2} c_{s}^{5}\:\gamma^{2}\:\eta^4\:k^5}.
\end{equation}

Once more, we see that the result does not depend on the collapse
times, removing the possibility of adjusting the dependence of
$\eta^{k}_{c}$ with $k$, in order to recover the scale free
spectrum. Also noteworthy is the fact that the above result is
independent of the parameter $l_0$ characterizing the birth time
of the modes.

\section{Conclusions}

In \cite{wald}, Hollands and Wald presented a cosmological scheme
that leads to similar predictions regarding the spectrum of
primordial cosmological fluctuations as those emerging from
inflationary models. The proposal is based on a model with a
matter content represented by more or less conventional fluids.
They studied a quantum treatment of the quantity representing the
coupled gravity-matter modes of the perturbations, and showed
that, under some simple hypotheses concerning the birth time and
the initial state of the relevant modes, it is possible to obtain
the same density perturbation spectrum and amplitude, as those
emerging from the successful inflationary models.

However, this model (just as the standard inflationary models),
has a serious issue regarding the lack of a satisfactory account
for the breaking of the symmetry of the initial state. Very
briefly, the issue is connected with interpretational difficulties
of quantum theory that become extremely exacerbated in the
cosmological setting. As has been discussed elsewhere, a
satisfactory model should explain the emergence of the seeds of
cosmic structure, which clearly correspond to inhomogeneities and
anisotropies. Starting from a perfectly symmetric quantum state
(isotropic and homogeneous), it cannot be accomplished in standard
terms given that the dynamics of the theory preserves such
symmetries. The Copenhagen approach to quantum theory cannot
explain how this transition would happen in the absence of
observers or external measurements.

In previous works, this issue has been extensively discussed, and
the arguments were not repeated here, so the interested reader is
referred to, for instance, \cite{sud, pss}.

In this manuscript, we showed that, in the framework of the model
considered in \cite{wald}, and with the same hypotheses, one can
obtain under certain circumstances, a prediction for an almost
scale free spectrum for scalar perturbations, when a collapse
scheme is introduced to allow for the transition from the original
state to a state containing actual seeds of cosmic structure. We
also showed which would be the dominant corrections that can be
expected to arise as a result of such a collapse.

One interesting feature that we found is that, under the
appropriate conditions, the amplitude of the dominant term leading
to the standard flat spectrum (corresponding to the term
proportional to $1/k^3$) does not depend on the collapse time.
Thus just as in \cite{wald}, the value of observed amplitude
($\sim 10^{-10}$) can be obtained simply by taking
$l_0\simeq5\times10^{-55}\:\rm{Mpc}$.

In the original proposal \cite{wald}, it was found that for the
case of pure radiation ($w=1/3$), the scale free spectrum cannot
be recovered. We considered the possibility that the introduction
of the collapse hypothesis might lead to a reversal of that
conclusion. However, we found that the final result for the
spectral shape is again independent of the collapse times.
Therefore, such a possibility is excluded, at least for the kind
of collapse models explored here.

Furthermore, our result (\ref{diezsiete}) also illustrates the fact
that corrections to the scale free spectrum can be expected in
association with the introduction of collapse models into the
theoretical analysis.

Finally, we should mention that we have focused only on the observational constraints associated with the amplitude and zeroth order form of the power spectrum. We view these as the most robust aspects of the analysis and the ones that are more likely to be roughly independent of the collapse scheme (i.e. the recipe for the selection of expectation values of the field and momentum conjugate in each mode, and the number of collapses per mode). However, other interesting
observables do exist, such as the tensor-to-scalar ratio and the scalar spectral index $n_S$. Regarding the latter, it is worthwhile noting something from Eq. (\ref{correciones}). In particular, if the collapse is assumed to occur at some physical fixed scale $l_c$, i.e. if $a(\eta_c^k)=k \:l_c$ (which is analogous to the  assumption for the 'birth time' of each mode), then the first corrective term in (\ref{correciones}) is scale invariant, and the second term is proportional to $k^{-(1/2+3 w/2)}$ indicating a red spectral index, which is what is favored by current observations. The detailed investigation of this issue is also left for future work. We must acknowledge that the model proposed in \cite{wald} leaves various issues unsolved, and it is not completely clear at this point if one can consider it as a truly viable alternative to inflation. However, our main point is that, just as the traditional inflationary paradigm, it requires some mechanism, such as the one explored in this work, to address the conceptual problem that afflict quantum schemes for the emergence of the seeds of cosmic structure, from homogeneous and isotropic initial states.

\bigskip

\acknowledgments{G.R.B. is supported by CONICET (Argentina).
G.R.B. acknowledges support from the PIP 2009-112-200901-00594 of
CONICET (Argentina), and he is grateful for the hospitality
received during his stay at the Instituto de Ciencias Nucleares
(UNAM), Mexico, where part of this work was done. We would also like to thank Gabriel Leon for his helpful and interesting
discussions. D.S.'s work  is supported in part by the CONACYT
grant No 220738  and  by  UNAM-PAPIIT  grant IN107412.}


\begin{thebibliography}{99}

\bibitem{WALD-QFTCS}R. M. Wald, Quantum field theory in curved space-time and black hole thermodynamics, 1992 (University of Chicago Press, Chicago,
USA).

\bibitem{H-Z} E. R. Harrison, Phys. Rev. D1, (1970) 2726; Y. B. Zel'dovich, Mon. Not. R. Astron. Soc. 160, (1972) 1p.

\bibitem{wmap9} G. Hinshaw et al., Astroph. J. Suppl. 208 (2013) 19.

\bibitem{planck1} Planck Collaboration, Paper XVI, arXiv:1303.5076, 2013.

\bibitem{wald} S. Hollands and R. M. Wald, Gen. Rel. Grav. 34, (2002)
2043.

\bibitem{muk} V. F. Mukhanov et al., Phys. Rep. 215, (1992) 203.

\bibitem{sud} D. Sudarsky, Int. Journal Mod. Phys. D20, (2011) 509.

\bibitem{kiefer} C. Kiefer and D. Polarsky, Adv. Sci. Lett. 2, (2009)
164; J. J. Halliwell, Phys. Rev. D39, (1989) 2912; C. Kiefer, Nuc.
Phys. B, Proc. Suppl. 88, (2000) 255; D. Polarsky and A. A.
Starobinsky, Class. Quant. Grav. 13, (1996) 377.

\bibitem{weinberg} S. Weinberg, Cosmology, 2008 (Oxford University Press, USA), section 10.1,
pp. 470-485.

\bibitem{mukbook} V. F. Mukhanov, Physical foundations of
cosmology, 2005 (Cambridge University Press, UK), section 8.3.3,
pp. 340-348.

\bibitem{lyth} D. H Lyth and A. R. Liddle, The primordial density
perturbation: cosmology, inflation and the origin of structure,
2009 (Cambridge University Press, UK), section 24.2, pp. 386.

\bibitem{penbook} R. Penrose, The road to reality: a complete
guide to the laws of the universe, 2004 (Vintage Books, USA),
section 30.14, pp. 861-865.

\bibitem{pss} A. Perez, H. Sahlmann and D. Sudarsky, Class. Quant. Grav.
23, (2006) 2317.

\bibitem{Bhom-Bub} D. Bohm and J. Bub, Rev. Mod. Phys. 38, (1966) 453.

\bibitem{Pearle:76} P. Pearle, Phys. Rev. D13, (1976) 857.

\bibitem{Pearle:79} P. Pearle, Int. J. Theor. Phys. 18, (1979) 489.

\bibitem{PCS:13} P. Ca\~nate, P. Pearle and D. Sudarsky, Phys. Rev. D87, (2013) 104024.

\bibitem{GRW:85} G. Ghirardi, A. Rimini and T. Weber, A model for a unified quantum description of macroscopic and
microscopic systems, in A. L. Accardi (Ed.) Quantum Probability
and Applications, 1985 (Springer, Heidelberg), pp. 223-232.

\bibitem{GRW:86} G. Ghirardi, A. Rimini and T. Weber, Phys. Rev. D34, (1986) 470.

\bibitem{Pearle:89} P. Pearle, Phys. Rev. A39, (1989) 2277.

\bibitem{GRW:90} G. Ghirardi, P. Pearle and A. Rimini, Phys. Rev. A42, (1990) 7889.

\bibitem{jm} J. Martin, V. Vennin and P. Peter, Phys. Rev. D86, (2012) 103524.

\bibitem{das} S. Das et al., Phys. Rev. D88, (2013) 085020; Erratum Phys. Rev. D89, (2014) 109902.

\bibitem{penrose1} R. Penrose, The emperor's new mind. Concerning
computers, minds and laws of Physics, 1989 (Oxford University
Press, UK).

\bibitem{penrose2} R. Penrose, in Physics meets Philosophy at the Planck
scale: Contemporary Theories in Quantum Gravity, (2011), edited by
C. Callendar and N. Huggett .

\bibitem{penrose3} R. Penrose, Gen. Rel. Grav. 28, (1996) 581.

\bibitem{diosi0} L. Diosi, Phys. Lett. A105, (1984) 199.

\bibitem{diosi1} L. Diosi, Phys. Lett. A120, (1987) 377.

\bibitem{diosi2} L. Diosi, Phys. Rev. A40, (1989) 1165.

\bibitem{Bhomian} A. Valentini, Phys. Rev. D82, (2010) 063513; N. Pinto-Neto, G. Santos and W. Struyve, Phys. Rev. D85, (2012) 083506.

\bibitem{gabriel} A. Diez-Tejedor et al., Gen. Rel. Grav. 44, (2012) 2965.

\bibitem{landau} S. J. Landau, C. G. Scoccola and D. Sudarsky, Phys. Rev. D 85, (2012) 123001.

\bibitem{mukv} V. F. Mukhanov, JETP Lett. 41, (1985) 493.

\bibitem{conceptual} R. Penrose, The Road to Reality, 2006 (Alfred A. Kopf- New York, USA); B. D. Espagnat, Phys. Lett. A282, (2001) 133; See for instance discussions about the EPR experiment in A.
Peres, Quantum Theory: Concepts and Methods, 1993 (Kluwer Academic
Publishers); D. Mermin, Physics Today 32, (1985) 38; M. Tegmark,
Fortsch. Phys. 46, (1998) 855, arXiv:quant-ph/9709032; A. Kent,
Int. J. Mod. Phys. A5, (1990) 1745; T. Sakaguchi, arXiv:
gr-qc/9704039, 1997; J. Garriga and A. Vilenkin, Phys. Rev. D64,
(2001) 043511; C. Barrabes and V. P. Frolov, Phys. Rev. D53,
(1996) 3215.

\bibitem{bassi} A. Bassi and G. C. Ghirardi, Phys. Rep. 379,
(2003) 257.

\bibitem{stat} G. Leon, S. J. Landau and D. Sudarsky, Phys. Rev. D88 (2013) 023526.

\bibitem{alberto} A. Diez-Tejedor and D. Sudarsky, JCAP 045,
(2012) 1207.

\bibitem{adolfo} A. de Unanue and D. Sudarsky, Phys. Rev. D78, (2008) 043510.

\bibitem{multiples} G. Leon, A. de Unanue and D. Sudarsky, Class. Quant. Grav. 28, (2011) 155010.


\end{thebibliography}
\end{document}